\documentclass[aps,prb,preprint,superscriptaddress]{revtex4-1}
\usepackage{graphicx}

\begin{document}


\title{Energetic disorder induced leakage current in organic bulk heterojunction solar cells: comprehending the ultra-high open circuit voltage loss at low temperatures}

\author{Wenchao Yang}
\author{Yongsong Luo}
\author{Pengfei Guo}
\author{Haibin Sun}
\affiliation{School of Physics and Electronic Engineering, Xinyang Normal University, Xinyang, 464000, China}
\author{Yao Yao}
\email{yaoyao@fudan.edu.cn}
\affiliation{Department of Physics, South China University of Technology, Guangzhou 510640, China}

\date{\today}

\begin{abstract}
In organic bulk heterojunction solar cells, the open circuit voltage ($V_\mathrm{oc}$) suffers from an ultra-high loss at low temperatures. In this work we investigate the origin of the loss through calculating the $V_\mathrm{oc}-T$ plots with the device model method systematically and comparing it with experimentally observed ones. When the energetic disorder is incorporated into the model by considering the disorder-suppressed and temperature-dependent charge carrier mobilities, it is found that for nonselective contacts the $V_\mathrm{oc}$ reduces drastically under the low temperature regime, while for selective contacts the $V_\mathrm{oc}$ keeps increasing with the decreasing temperature. The main reason is revealed that as the temperature decreases, the reduced mobilities give rise to low charge extraction efficiency and small bimolecular recombination rate for the photogenerated charge carriers, so that in the former case they can be extracted from the wrong electrode to form a leakage current which counteracts the photocurrent and increases quickly with voltage, leading to the anomalous reduction of $V_\mathrm{oc}$. In addition, it is revealed that the charge generation rate is slow-varying with temperature and does not induce significant $V_\mathrm{oc}$ loss. This work also provides a comprehensive picture for the $V_\mathrm{oc}$ behavior under varying device working conditions.
\end{abstract}


\maketitle

\section{Introduction}
For the state-of-the-art organic bulk heterojunction solar cells, the power conversion efficiency has exceeded 10\% through decades of efforts on the optimization of device structures and internal morphology\cite{deibel}. Nevertheless, they are far from commercial applications due to the ubiquitously low open circuit voltages ($V_\mathrm{oc}$) of no more than 1\,V, which implies a remarkable loss when comparing the $eV_\mathrm{oc}$ with the effective band gap, namely the energy offset between the LUMO (Lowest unoccupied molecular orbital) of acceptor and the HOMO (highest unoccupied molecular orbital) of donor\cite{yao}. In order to reveal the origin of this ultra-high $V_\mathrm{oc}$ loss and find means to minimize it, people have investigated extensively the relationships between the $V_\mathrm{oc}$ and various types electronic properties or processes, such as the the charge transfer (CT) state energy\cite{deibel2,nmat,vandewal,burke,zou,guan}, the charge recombination dynamics\cite{maurano,oskar2}, the charge carrier densities\cite{rauh}, the energetic disorder\cite{neher,bel,belmonte,collins}, the light intensity\cite{koster,brus}, the internal molecular morphology\cite{perez}, the metal/organic contact properties\cite{oskar2,oskar1,mehai,me}, and so on. The results suggest that the $V_\mathrm{oc}$ is intricately correlated with all of these factors, and an elaborate theoretical formulation of $V_\mathrm{oc}$ is highly desirable\cite{uddin}.

There are generally two fundamental principles from which the basic expressions of $V_\mathrm{oc}$ and its losses could be derived. First, electrically, the $V_\mathrm{oc}$ is defined as the splitting between the quasi-Fermi level of electrons and that of holes under the open circuit condition, namely the non-equilibrium steady state in which the charge generation rate $G$ exactly equals to the charge recombination rate $R$\cite{neher}. Second, optically, there exists a detailed balance relation between the reversal processes of optical absorption and electroluminescence (radiative recombination) in a working solar cell\cite{rau,nmat,vandewal}. The two formulations give rise to the same behavior of $V_\mathrm{oc}$ following the variation of temperature, that is, the $V_\mathrm{oc}$ increases linearly with the decreasing temperature until saturating to its maximum value at the low temperature regime, which is experimentally demonstrated very well\cite{rand,hormann}. Employing this behavior people can obtain the diode ideality factors of organic photovoltaic devices under illuminated conditions by measuring the slope of the $V_\mathrm{oc}-T$ plots. The most significant distinction between the two formulations lies on the fact that they suggests two different maximum achievable $V_\mathrm{oc}$'s. According to the first principle the maximum $eV_\mathrm{oc}$ should be the effective band gap $E_\mathrm{g}$\cite{rauh,rand}, while following the second one it is the CT state energy $E_\mathrm{CT}$\cite{deibel2,hormann,zou,guan}. When considering the energetic disorder of the LUMO/HOMO levels, or that of the CT state manifold ubiquitous at the organic donor/acceptor heterojunction, the density of states (DOS) of the relevant electronic states are broadened and extended into the band gap, and the $V_\mathrm{oc}$ expression should be corrected due to the relaxation of charge carriers to the tail states or the CT excitons to the low energy vibronic states\cite{grancini,zhu}. For the first case, under the assumption of Gaussian disorder model, the total $V_\mathrm{oc}$ expression can be written as
\begin{equation}\label{voc}
  eV_\mathrm{oc}=E_\mathrm{g}-\frac{\sigma_n^2+\sigma_p^2}{2kT}-kT\ln\left(\frac{N_cN_v}{np}\right),
\end{equation}
where the $\sigma_n(\sigma_p)$ are the widths of the LUMO (HOMO) DOS, and the other notations follow the common definitions\cite{neher,bel,belmonte}. For the second case, we just need to replace the second term on the right side of Eq.(\ref{voc}) with $\sigma_\mathrm{CT}^2/2kT$, in which $\sigma_\mathrm{CT}$ is the DOS width of the CT state manifold\cite{burke,liu}. Thus the $V\mathrm{oc}-T$ relations deduced from the two principles remain follow the same behavior even though the respective physical meanings of disorder may be different. Around the room temperature, the $V_\mathrm{oc}-T$ relation is still dominated by the linear increasing behavior as the temperature reduces. Whereas under the low temperature regime, the correction term which is inversely proportional to the temperature becomes innegligible. Except from reducing the maximum achievable $V_\mathrm{oc}$, It makes the $V_\mathrm{oc}$ increase gradually and saturate slower with the decreasing temperature.

Recently, Gao et al. observed a novel phenomenon that in certain types of polymer-fullerene blended solar cells, as the temperature decreases from around the room temperature (RT), the $V_\mathrm{oc}$ initially increases linearly as expected, but it begins to decrease below a particular low temperature\cite{gao}. To our knowledge, this non-monotonic variation of $V_\mathrm{oc}-T$ relation had never been reported before, and its underlying mechanism needs to be elucidated. The phenomenon is attributed to the simultaneous deceasing of the photo-generated charge carrier density because of the diminishing of entropy-driven charge separation process at low temperatures\cite{breg,xyz}, which could lead to the reduced $V_\mathrm{oc}$ according to the third term on the right hand side of Eq.~(\ref{voc}). But this has not been verified quantitatively through device simulation works. On the other hand, it can also be expected that at low temperatures the energetic disorder may suppress the linear increasing behavior of the $V_\mathrm{oc}$ as stated above. However, with realistic energetic disorder parameters substituted into Eq.~(\ref{voc}), the calculated $V_\mathrm{oc}-T$ relation still gives rise to monotonically increasing behavior with decreasing temperature\cite{bel}. In this work we endeavor to find the origin of this anomalous reduction of $V_\mathrm{oc}$ through the calculations of the $J-V$ curves and the $V_\mathrm{oc}-T$ relations for various device parameters and working conditions. In particular, we incorporate the disorder-suppressed charge carrier mobility obtained through the kinetic Monte Carlo simulations for organic solid into the device model, which is found to be the key factor for inducing the reduction of the $V_\mathrm{oc}$. The results are helpful for a comprehensive understanding of the temperature dependence of the $V_\mathrm{oc}$ and the underlying mechanisms through which the disorder may damage the performance of organic bulk heterojunction solar cells.

\section{Device model method}
The device model (drift-diffusion) method is commonly employed for simulating the macroscopic device operating properties. In this model, the photoactive layer is basically considered as a homogeneous medium. Then the time evolution of the photogenerated CT states and the charge carriers are described by the following one-dimensional continuity equations,
\begin{eqnarray}
  \frac{\partial X}{\partial t} &=& D_X\frac{\partial^2}{\partial x^2}-k_dX-\frac{X}{\tau}+G, \label{xt} \\
  \frac{\partial p}{\partial t} &=& -\frac{1}{e}\frac{\partial J_p}{\partial x}+k_dX-R(x), \label{pt} \\
  \frac{\partial n}{\partial t} &=& \frac{1}{e}\frac{\partial J_n}{\partial x}+k_dX-R(x).  \label{nt}
\end{eqnarray}
$X, p, n$ denote the densities of the CT states, the holes and the electrons, respectively. On the right hand side of Eq.~(\ref{xt}), the first term stands for the CT states diffusion (with $D_X$ the diffusivity), the second one for the CT state dissociation and charge formation (with $k_d$ the dissociation rate), the third one for the radiative and nonradiative decay of CT states to the ground state (with $\tau$ the lifetime), and the fourth term for the photogeneration of the CT states (with G the generation rate). The hole (electron) current density in Eq.~(\ref{pt})(Eq.~(\ref{nt})) is of the common drift-diffusion form:
\begin{equation}\label{dd}
  J_{p(n)}(x)=e\mu_p\left(p(n)F\mp \frac{kT}{e}\frac{\partial p(n)}{\partial x}\right).
\end{equation}
The spatio-temporal evolution of the internal electric field $F$ is governed by the Poisson's equation, which reads
\begin{equation}\label{possion}
  \frac{\partial F}{\partial x}=\frac{e}{\epsilon_0\epsilon}(p-n).
\end{equation}
By solving Eqs.~(\ref{xt},\ref{pt},\ref{nt},\ref{possion}) numerically many types of macroscopic quantities for a working device can be calculated. The details for device model implementation are elaborated in numerous literatures and thus are omitted here\cite{smith,blom,finck}. In our simulation, initially the Onsager-Braun theory is incorporate into the model to account for the temperature variation of the CT state dissociation (or charge generation) rate, which can be written as\cite{blom,clark}
\begin{equation}\label{kd}
  k_d=\frac{3\gamma}{4\pi a^3}\exp\left(-\frac{E_b}{kT}\right)\left(1+b+\frac{b^2}{3}+\frac{b^3}{18}\right),
\end{equation}
where the parameter
\begin{equation}\label{bf}
  b=\frac{e^3F}{8\pi \epsilon_0\epsilon k^2T^2 },
\end{equation}
and
\begin{equation}\label{rf}
  \gamma=\frac{e(\mu_n+\mu_p)}{\varepsilon_0\varepsilon}
\end{equation}
is the Langevin recombination coefficient. $E_b=e^2/4\pi \varepsilon_0\varepsilon a$ is the CT state binding energy with $a$ the CT state radius. The other notations follow the common definitions. For the recombination rate $R$, the Langevin-type bimolecular recombination is assumed to be the dominant recombination mechanism, namely
\begin{equation}\label{rr}
  R=\gamma(np-n_i^2).
\end{equation}

When the effects of the intrinsic energetic disorder are to be investigated, instead of analyzing the effective reduction of the quasi-Fermi level splitting\cite{bel,belmonte}, we incorporate it into the device model through considering its influence on the charge transport properties. With the presence of the disorder, the charge carriers undergo the phonon-assisted hopping motion due to the induced charge localization effect. By employing semiclassical numerical methods like the kinetic Monte Carlo simulation (KMC), the non-equilibrium steady state charge mobilities $\mu$ were calculated many times for the Gaussian disorder model, and were found to be of the following temperature dependence\cite{bassler,cordes,pasveer,ctr,kuik}:
\begin{equation}\label{mu}
  \mu=\mu_{\infty}\exp\left[-\left(c\frac{\sigma}{kT}\right)^2\right],
\end{equation}
where $\mu_{\infty}$ is the high temperature limit of the mobility, $\sigma$ is the Gaussian disorder width and the coefficient $c$ is set to be 2/3\cite{deibel}. According to Eq.~(\ref{mu}) the mobility is strongly suppressed by the energetic disorder under low temperatures. In additions, the same calculations also revealed that the energetic disorder can make the charge mobility become dependent on the charge density and the electric field, but these two dependencies only become significant when the carrier density and the electric field strength are sufficiently high such as those in organic field effect transistors. However in organic solar cells these effects are negligible, and for simplicity they are not included in our model. It should be noted that if the temperature dependent mobility (Eq.~(\ref{mu})) is involved when calculating the steady state quantities, with decreasing temperature the evolution time required for the device to reach the steady state becomes longer and thus need to be enhanced accordingly. The simulation parameters are listed in table (\ref{parameter}) except noted otherwise.

\begin{table}
 \caption{The parameters used in the device model simulation\label{parameter}}
 \begin{ruledtabular}
 \begin{tabular}{ccc}
 Parameter &  Symbol & Value \\
  \hline
  Effective band gap & $E_g$ & 1.3\,eV   \\
  Injection barriers & $\phi_n, \phi_p$ & 0.2\,eV \\
  Relative permitivity & $\varepsilon$ & 3.5  \\
  Active layer thickness & $L$   & 200\,nm \\
  Density of states & $N_C, N_V$  & $10^{21} \mbox{cm}^{-3}$ \\
  High-$T$ Mobility & $\mu_{\infty}$ & $10^{-5}\,\mbox{m}^2/\mbox{Vs}$ \\
  CT generation rate & $G$  & $3\times 10^{21} \mbox{cm}^{-3}\mbox{s}^{-1}$ \\
  CT state lifetime & $\tau$ & 100\,ns \\
  CT state radius &  $a$  & 2\,nm \\
 \end{tabular}
 \end{ruledtabular}
\end{table}

\section{Results and discussion}
\begin{figure}
  \centering
  \includegraphics[width=12cm]{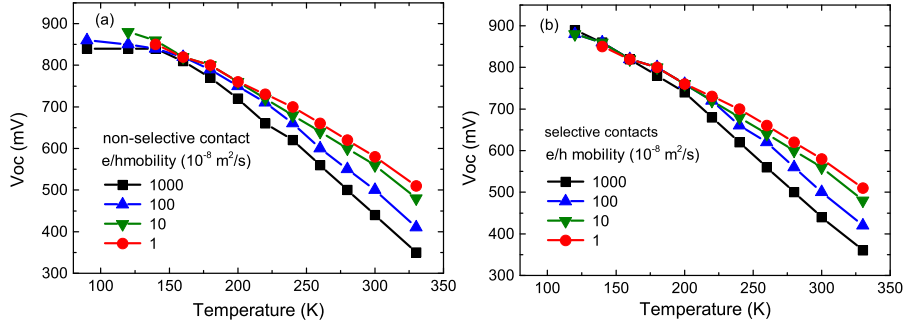}
  \caption{(a) The $V_\mathrm{oc}-T$ plots calculated under a set of balanced electron and hole mobilities with nonselective contacts. (b) The $V_\mathrm{oc}-T$ plots calculated under the same condition except with selective contacts being assumed. The Onsager-Braun type CT state dissociation rate $k_d$ is incorporated in the calculation of both of the plots. }\label{contact}
\end{figure}

First of all, we attempt to verify the hypothesis that the decreased photo-generated charge carrier density causes the reduction of $V_\mathrm{oc}$ at low temperatures. Meanwhile, the influence of the metal/organic interfacial properties on $V_\mathrm{oc}$ are also examined by assuming different boundary conditions in the device model. With constant and balanced electron/hole mobilities being substituted into the model, we calculate the $J-V$ curves for a set of temperatures ranging from 320\,K down to ones at which the photo-generated charge carriers are very limited and the photocurrent decreases approximately to zero, for instance 100\,K. From the $J-V$ curves the $V_\mathrm{oc}$'s corresponding to each temperature can be extracted. In Fig.~\ref{contact} the $V_\mathrm{oc}-T$ curves with nonselective and selective contacts for charge carrier extraction are shown, respectively. It is found that as temperature decreases from above the room temperature (RT), the linearly increasing behavior of $V_\mathrm{oc}$ are clearly reproduced for both of the two types of contacts. With nonselective contact, when the temperature reduces to 150\,K the $V_\mathrm{oc}$'s gradually saturate to below 900\,mV, which is the the maximum achievable photovoltage given by the band gap minus the electron and hole injection barriers. For higher carrier mobility such as $10^{-5}\,\mbox{m}^2/\mbox{Vs}$, the calculated $V_\mathrm{oc}$ is relatively lower and saturates earlier with the decreasing temperature, since high mobility may brings about large bimolecular recombination rate (see Eq.~(\ref{rf})) and severe surface losses due to the extraction of charge carriers from the wrong electrode. With selective contact, the $V_\mathrm{oc}-T$ plots for different mobilities converge below 180\,K. They exhibit no obvious saturation behavior but finally increase to a value that is very close to 900\,mV, because the surface losses are absent in this case. With these conditions, no $V_\mathrm{oc}$ reduction at low temperatures is observed, which suggests that it is not sufficient to explain this phenomenon solely by the reduction of charge carrier density with the decreasing temperature due to the Onsager-Braun theory or entropy effect, and for the constant mobility cases the metal/organic contact properties have little influence on the $V_\mathrm{oc}$ behaviors. In the following, except stated explicitly all the simulations are done with non-selective contacts.

\begin{figure}
  \centering
  \includegraphics[width=12cm]{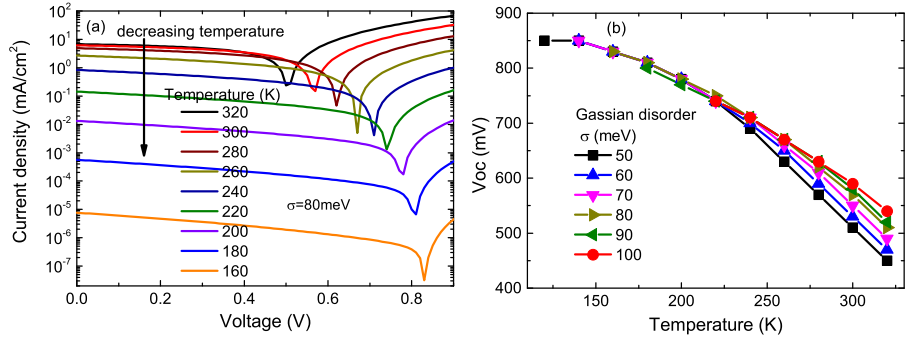}
  \caption{(a) The $J-V$ curves calculated under different temperatures for disorder suppressed mobilities, in which the energetic disorder $\sigma=$\,80meV. (b) The $V_\mathrm{oc}-T$ plots with the disorder suppressed carrier mobilities. The Onsager-Braun type CT state dissociation rate $k_d$ is incorporated in the calculation of both of the plots.  }\label{jvvoc}
\end{figure}

Now we examine the influenced of the energetic disorder on the $V_\mathrm{oc}-T$ relations. Incorporating Eq.~(\ref{mu}) and the Onsager-Braun charge generation rate of Eq.~(\ref{kd}) into the device model, the $J-V$ curves for different temperatures are calculated and presented in Fig.~\ref{jvvoc}(a), in which the the Gaussian disorder $\sigma$ is set to be 80\,meV. It is observed that as the temperature decreases the reduction of the short circuit current density $J_\mathrm{sc}$ becomes more and more drastic because of the quick reductions of both the charge carrier generation rate and the mobilities, while the $V_\mathrm{oc}$ increases more and more slowly under the low temperature regime but exhibits no reduction behavior. In Fig.~\ref{jvvoc}(b) the calculated $V_\mathrm{oc}-T$ plots for a wide range of typical Gaussian disorder are shown. For small disorder such as $\sigma=50\,\mbox{meV}$, the linear increasing behavior is still retained for a relatively large temperature range before the final saturation; whereas for the energetic disorder like $\sigma=100\,\mbox{meV}$, although the $V_\mathrm{oc}$ at RT becomes slightly higher, the increasing of $V_\mathrm{oc}$ becomes slower (see the slope change of the $V_\mathrm{oc}-T$ plots for different disorder parameters at the high temperature regime) and quickly deviates from the linear increasing behavior with decreasing temperature. At the low temperature regime, the $V_\mathrm{oc}-T$ plot with $\sigma=100\,\mbox{meV}$ tends to saturate at a rather small value less than 800\,mV. This disorder-induced $V_\mathrm{oc}$ loss is purely originated from the disorder induced mobility reduction, rather than the diminishing of the effective band gap due to the charge carrier relaxation. Nevertheless, even for such a high energetic disorder as 100\,meV the corresponding $V_\mathrm{oc}-T$ plot still increases monotonically with decreasing temperature. Therefore the combined effect of the strongly temperature-dependent charge generation rate and the disorder-suppressed mobilities do not lead to the reduced $V_\mathrm{oc}$ at low temperatures.

\begin{figure}
  \centering
  \includegraphics[width=12cm]{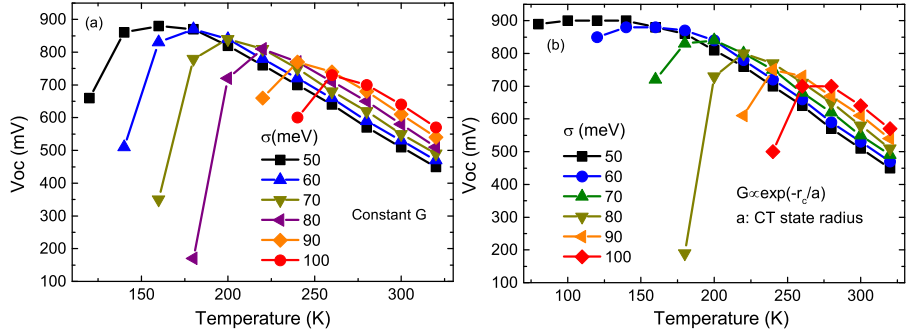}
  \caption{(a) The calculated $V_\mathrm{oc}-T$ plots for fully constant charge generation rate and disorder suppressed mobilities, with different Gaussian disorder values being assumed. (b) The calculated $V_\mathrm{oc}-T$ plots under the same conditions as (a) except that a slow-varying charge generation rate are assumed.}\label{vocr}
\end{figure}

We consider the sole effect of disorder-suppressed mobilities on the $V_\mathrm{oc}$ loss. By retaining Eq.~(\ref{mu}) for the disorder-suppressed mobilities but using a fully constant charge carrier generation rate $G$ in the device model, the $V_\mathrm{oc}-T$ plots are calculated and shown in Fig.~\ref{vocr}(a). It is found that the linear increasing behavior just below the RT is still present. More importantly, the reduction of $V_\mathrm{oc}$ indeed emerges for each of the Gaussian disorder $\sigma$ when the temperature decreases below a critical point. With the small $\sigma$ of 50\,meV, the $V_\mathrm{oc}$ decreases if the temperature is lower than 160\,K. As the disorder increases, the critical temperature corresponding to the maximum $V_\mathrm{oc}$ increases monotonically. For $\sigma=100$\,meV, the critical temperature falls on 260\,K and the maximum $V_\mathrm{oc}$ is only 750\,mV. Compared with the experimentally measured $V_\mathrm{oc}-T$ plots by Gao et al., the simulated decreasing behavior is much more drastic. This effect may be derived from that in our simulations the strongly temperature-dependent mobilities are used, while in real systems the temperature variation of mobilities could be much milder than that described by Eq.~(\ref{mu}). The successful reproduction of the experimentally observed $V_\mathrm{oc}$ reduction in Fig.~\ref{vocr}(a) also suggests that the Onsager-Braun theory may greatly overestimate the impact of temperature on CT states dissociation and charge generation. Actually this is consistent with the findings that a high percentage of charge carriers are generated through the quantum coherence dominated ultrafast charge transfer processes at the donor/acceptor interfaces\cite{kaake,whaley}, which is expected to be more prominent at low temperatures. On the other hand, it is unrealistic to assume a temperature-independent charge carrier generation rate as well. Therefore we set the coefficient $\gamma$ (Eq.~(\ref{rf})) in the Eq.~(\ref{kd}) to be disorder-free, and the temperature-dependence of $k_d$ is predominantly described by the Coulomb capture radius $r_c$, which is defined as $r_c=e^2/4\pi\varepsilon_0\varepsilon kT$, and appears in the factor of $\exp(-r_c/a)=\exp(-E_b/kT)$. The calculated $V_\mathrm{oc}-T$ plots for the $k_d$'s with different temperatures are shown in Fig.~\ref{vocr}(b). It can be seen that the main features of these plots are still the same as those shown in Fig.~\ref{vocr}(a), except that for small $\sigma$'s the decreasing behavior of the $V_\mathrm{oc}$ becomes moderate at the low temperature regime, which more closely resembles the experimentally measured $V_\mathrm{oc}-T$ plots. In addition, at the RT regime the higher the disorder, the larger is the $V_\mathrm{oc}$, because the the high-disorder suppressed mobilities induce low bimolecular recombination rate $R$ and thus small internal losses. According to these results, we can conclude that it is the disorder suppressed charge carrier mobility that plays the utmost important role on inducing the ultra-high $V_\mathrm{oc}$ loss with decreasing temperature.

\begin{figure}
  \centering
  \includegraphics[width=8cm]{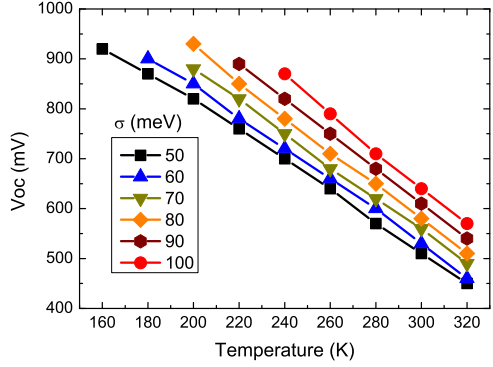}
  \caption{The calculated $V_\mathrm{oc}-T$ plots for different Gaussian disorder $\sigma$'s. The selective contacts are assumed at both the anode/organic and organic/cathode interfaces, and the CT states dissociation rate $k_\mathrm{d}$ is only proportional to $\exp(-r_C/a)$. Under the large $\sigma$'s and low temperatures, the correct steady state could hardly be reached through time evolution in our program, so that more corresponding $V_\mathrm{oc}$ points down to low temperatures are not included in the plots.}\label{selective}
\end{figure}

\begin{figure}
  \centering
  \includegraphics[width=12cm]{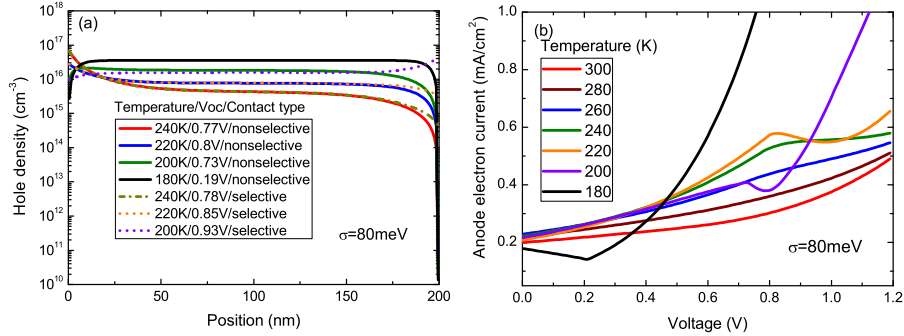}
  \caption{(a) The calculated hole density profiles under the open circuit conditions of different temperatures and contact types. The Gaussian disorder $\sigma$ is set to be 80\,meV. (b) The calculated $J-V$ curves of the electron current extracted from the anode with non-selective contacts under different temperatures.}\label{holedf}
\end{figure}

In Fig.~\ref{selective} we present the calculated $V_\mathrm{oc}-T$ plots with selective contacts while keeping other simulation conditions the same as above. It can be observed that even though under high Gaussian disorder values the $V_\mathrm{oc}$'s increases monotonically and linearly with approximately constant slope as the temperature reduces, in contrast to the abrupt reduction behavior with nonselective contacts. Furthermore, the $V_\mathrm{oc}$ is not restricted by the injection barriers at the contacts, so that it can exceed 900\,mV under low temperature. This suggests the charge extraction at the wrong electrodes, or surface losses plays significant roles on lowering the $V_\mathrm{oc}$ with decreasing temperature. To reveal the underlying mechanism, we calculated the hole density profile under the open circuit conditions for both selective (dashed lines) and non-selective (solid lines) contacts, with a Gaussian disorder of 80\,meV, as shown in Fig.~\ref{holedf}(a). Comparing the hole density profiles for the two contact types at the same temperature, it is apparent that they precisely overlap together in the bulk of the device but diverge from each other at the vicinity of the contacts. This divergence becomes significant as the temperature decreases. At the relatively high temperature of 240\,K, the overlapped region extends to the anode side where the equilibrium hole density is much higher than those in the bulk, and the electrons moving to the anode can be effectively annihilated through intensive bimolecular recombination, such that the nonselective contacts are essentially selective at high temperatures. Whereas at low temperatures, although the bulk hole density is increased as a result of the quick reduction of the bimolecular recombination coefficient $\gamma$ which involves the disorder-suppressed mobility $\mu$ (Eq.(\ref{rf})), the interfacial equilibrium hole density is significantly decreased as is given by the Boltzmann factor $\exp(-\phi/kT)$ (with $\phi$ being the injection barrier), and the density can even be lower than that in the bulk. Therefore for nonselective contacts, the electrons in the bulk may enter the anode without any kind of blocking effect, which brings about significant surface losses.

This loss mechanism can be manifested by the calculated $J-V$ curves of electron current extracted to the anode under various temperatures for the nonselective contact case, as shown in Fig.~\ref{holedf}(b). Under the high temperatures, due to the blocking effect induced by the high equilibrium hole density near the contact, the interfacial electron current is small and increases very slowly with bias voltages. As the temperature decreases, this effect vanishes gradually and the electron current increases. When the temperature reduces to 200\,K or below, plenty of electrons are accumulated in the bulk region because of the greatly reduced bimolecular recombination rate, which can be easily extracted to the anode when the external bias voltage increases. Therefore in this case, the reduced mobility causes the anode electron current to decrease at the short circuit condition, but this current component (acting as the leakage current) increases extremely quickly with increasing voltage and counteracts with the reduced hole current, making the direction of the total photocurrent reverse under a small bias voltage, which leads to the reduced $V_\mathrm{oc}$ with nonselective contacts. The same argument can be applied to the hole current extracted from the cathode, which also contributes to the large leakage current in the high voltage regime. Now it is understandable that why the $V_\mathrm{oc}$ does not decrease when the Onsager-Braun theory is assumed in the device model, because in this case the reduction of the charge carrier density with decreasing temperature is so drastic that there is very little free charge carriers in the bulk that would probably be extracted from the wrong electrode, and thus the undesirable leakage current is negligible. Inserting electron (hole) blocking layers between the anode (cathode) and the photoactive layer may effectively circumvent the drastic reduction of $V_\mathrm{oc}$, but the reduction of photocurrent due to the small disorder-suppressed charge carrier mobilities is inevitable.

\section{Conclusions}
In this work, we employed the one-dimensional device model simulation method to investigate the anomalous reduction of the open circuit voltage at low temperatures in organic solar cells. The $V_\mathrm{oc}-T$ plots are calculated under varied charge generation rate, charge carrier mobilities, metal/organic contact types and many other different conditions. It is found that with the strongly temperature-dependent Onsager-Braun type charge generation rate and the temperature-independent constant mobilities, the $V_\mathrm{oc}$ just follows the common behavior of linear increasing as the temperature decreases and finally saturate, so that the reduced charge carrier density under low temperature is not sufficient for explaining the anomalous reduction behavior. After incorporating the dual effects of Onsager-Braun type charge generation rate and the energetic disorder-suppressed mobilities derived from the Gaussian disorder model, the short circuit current is greatly reduced, but still the behavior of $V_\mathrm{oc}$ does not show any significant change. For the nonselective contacts, the low temperature reduction behavior of $V_\mathrm{oc}$ indeed emerges in the case of incorporating the disorder-suppressed mobilities while keeping the charge generation rate constant or only slow-varying with temperature (its temperature dependence is only contained in $r_\mathrm{C}$). The larger the Gaussian disorder $\sigma$, the higher is the temperature below which the $V_\mathrm{oc}$ begins to decrease. However, with selective contacts being assumed, the $V_\mathrm{oc}$'s do not decrease at low temperatures no matter how large is the $\sigma$'s. The results suggest the origin of the ultra-high $V_\mathrm{oc}$ loss for the nonselective contacts is that due to the decreasing of the interfacially accumulated majority charge carriers and the bimolecular recombination rate with the decreasing temperature, the minority charge carriers cannot be effectively blocked from entering the respective wrong electrodes, leading to high leakage current as the bias voltage increases, and the leakage currents counteract with the extracted photocurrent until the total current vanishes at a small voltage, which is exactly the reduced $V_\mathrm{oc}$. This mechanism is manifested by the calculated hole density profiles under the open circuit conditions and the temperature variation of the $J-V$ curves for the anode interfacial electron current with nonselective contacts.
\begin{acknowledgments}
The author acknowledges the financial support from the National Science Foundation of China under the grant No.11604280, and the starting research grant for the junior investigators of Xinyang Normal University.
\end{acknowledgments}

\bibliography{citation}

\end{document}